\documentclass[seceq]{jpsj2}

\newcommand{\rmi}{\textrm{i}}
\newcommand{\rmd}{\textrm{d}}
\newcommand{\rme}{\textrm{e}}
\newcommand{\kf}{k_{\textrm{F}}}
\newcommand{\hc}{h_{\textrm{c}}}
\newcommand{\calH}{\mathcal{H}}

\def\runtitle{Emptiness Formation Probability for the One-Dimensional Isotropic ${XY}$ Model}
\def\runauthor{Masahiro {\sc Shiroishi}, Minoru {\sc Takahashi} and Yoshihiro {\sc Nishiyama}}

\title
{
Emptiness Formation Probability for the One-Dimensional Isotropic ${XY}$ Model
}

\author
{ 
Masahiro {\sc Shiroishi}$^{1}$\thanks{E-mail address: siroisi@issp.u-tokyo.ac.jp}, Minoru {\sc Takahashi}$^{1}$\thanks{E-mail address: mtaka@issp.u-tokyo.ac.jp}
and Yoshihiro {\sc Nishiyama}$^{2}$\thanks{E-mail address: kitarou@soroban.phys.okayama-u.ac.jp}
}

\inst
{
$^1$Institute for Solid State Physics, University of Tokyo, \\ 
Kashiwanoha 5-1-5, Kashiwa, Chiba, 277-8571 \\
$^2$Department of Physics, Faculty of Science, Okayama University, \\ 
Okayama 700-8530}

\recdate
{
}

\abst
{
We study a correlation function for the one-dimensional isotropic 
${XY}$ model (${XX0}$ model), which is called the Emptiness Formation 
Probability (EFP).  It is the probability 
of the formation of a ferromagnetic string in the anti-ferromagnetic 
ground state.  Using the expression of the EFP as a Toeplitz determinant, 
we discuss its asymptotic behaviors.  
We also compare the analytical results with numerical calculations as the density-matrix renormalization group and the quantum Monte-Carlo method.
}

\kword
{
emptiness formation probability, isotropic XY model, DMRG method, finite temperature, QMC method
}

\begin{document}
\sloppy
\maketitle

\section{Introduction}
Recently there have been a lot of progresses for the calculation of correlation
functions for the one-dimensional ${XXZ}$ model \cite{Korepin93, Jimbo95, Korepin94, Essler95n2, Essler95, Kitanine00, Razmov00,Boos01},
\begin{equation}
\calH = J \sum^N_{j=1} \left\{ S^x_jS^x_{j+1}+S^y_jS^y_{j+1}+ \Delta
S^z_jS^z_{j+1} \right\} -2h\sum^N_{j=1}S^z_j,
\label{eq.XXZ}
\end{equation}
which is exactly solved by the Bethe ansatz method \cite{Takahashi99}.  Among various correlation functions, the so called {\it emptiness 
formation probability} (EFP) ${P(n)}$ 
\begin{equation}
P(n) = \langle \prod^n_{j=1}(S^z_j+\frac{1}{2}) \rangle,
\end{equation}
is the most fundamental one \cite{Korepin94}. 

Concerning the asymptotic behaviours of ${P(n)}$ for ${|\Delta| \le 1}$,
the following properties have been discussed: \\
(1)  ${P(n)}$ decays like a Gaussian at zero temperature \cite{Essler95}.  \\
(2)  ${P(n)}$ decays exponentially at finite temperature \cite{Boos01}. \\
In this paper, we shall confirm these properties for the isotropic 
${XY}$ model (${\Delta=0}$) \cite{Lieb61, Katsura62} in detail.
Actually, based on the representation of ${P(n)}$ as a Fredholm
determinant \cite{Essler95}, (c.f. eq. (\ref{eq.fredholm})), the property (1) has been established for generic ${|\Delta| \le1}$ to some extent 
.

For the special case, ${\Delta=0}$, we will show in this paper that ${P(n)}$ 
can be alternatively represented by a Toeplitz determinant. The representation 
by Toeplitz determinant is very powerful to give us more information on
the properties of ${P(n)}$. Actually there are theorems known for the
asymptotics of the Toeplitz determinants, which directly establish
properties (1) and (2). They also give us the complete forms of the
asymptotic behaviours including the prefactor. 

On the other hand, we can calculate ${P(n)}$ for small ${n}$ by other numerical methods.  In this paper we employ the density-matrix renormalization group (DMRG) method for zero temperature case and the quantum Monte-Carlo (QMC) method for finite temperature case. The numerical data calculated  by the DMRG and the QMC  agree satisfactorily with the analytical ones. We believe the method and the results presented in this paper will afford a good foundation when we investigate the EFP for other models.

The plan of the paper is as follows. In \S 2, we derive 
the representation of ${P(n)}$ by Toeplitz determinant. The asymptotic
form of ${P(n)}$ is obtained from Widom's theorem.
In \S 3, we demonstrate a numerical simulation of ${P(n)}$ by DMRG and compare with the analytical calculation. In \S 4, the EFP at the finite temperature is studied. This time the asymptotic function is obtained from Szeg\"{o}'s theorem. It is compared with the numerical calculations by QMC in \S 5.
The last section is devoted to summary and discussions.

\section{EFP at Zero Temperature}
The Hamiltonian of the isotropic ${XY}$ model is defined by the special
case of (\ref{eq.XXZ}) with ${\Delta=0}$, i.e.,
\begin{equation}
\calH = J \sum^N_{j=1} \left\{ S^x_jS^x_{j+1}+S^y_jS^y_{j+1} \right\} -2h\sum^N_{j=1}S^z_j, 
\label{eq.XY}
\end{equation}
where $S^\alpha_j$ are the spin 1/2 operators
\begin{equation}
[S^\alpha_k,S^\beta_l]=i\delta_{kl}\epsilon_{\alpha\beta\gamma}S^\gamma_k, \label{eq.spin}
\end{equation}
and are represented by the Pauli matrices $\sigma^\alpha_j$ as
\begin{align}
&S^\alpha_j=\frac{1}{2}\sigma^\alpha_j, \nonumber\\
&
\sigma^x_j=\left[\begin{array}{@{\,}cc@{\,}}
0 & 1 \\ 
1 & 0 \\ 
\end{array}\right], 
\sigma^y_j=\left[\begin{array}{@{\,}cc@{\,}}
0 & - \rmi \\ 
\rmi & 0 \\ 
\end{array}\right], 
\sigma^z_j=\left[\begin{array}{@{\,}cc@{\,}}
1 &  0 \\ 
0 & -1 \\ 
\end{array}\right]. 
\label{eq.spinoperator}
\end{align}
In the Hamiltonian (\ref{eq.XY}), we assume the periodic boundary 
condition (PBC),
\begin{equation}
S^\alpha_{N+1}\equiv S^\alpha_1. \label{eq.pbc}
\end{equation}
We also introduce ${S^\pm_k \equiv S^x_k\pm iS^y_k}$ for later convenience.
The EFP at zero temperature is defined by 
\begin{equation}
P(n) = \langle GS|\prod^n_{j=1}(S^z_j+ \frac{1}{2})|GS \rangle \label{eq.xyfep},
\end{equation}
which represents the probability of the formation of a ferromagnetic string 
with a length $n$ in the anti-ferromagnetic ground state ${|GS \rangle}$.

The Hamiltonian (\ref{eq.XY}) can be diagonalized by means of 
the Jordan-Wigner transformation \cite{Lieb61},
\begin{align}
S^-_k &= \prod^{k-1}_{j=1}(1-2c_j^\dagger c_j)c_k^\dagger,\nonumber\\
S^+_k &= \prod^{k-1}_{j=1}(1-2c_j^\dagger c_j)c_k, \label{eq.JW}
\end{align}
where $c^\dagger_k,c_k$ are the fermion creation and annihilation operators, 
respectively.  In fact, the Hamiltonian (\ref{eq.XY}) is transformed to 
a spinless fermion model as
\begin{align}
\calH &= -\frac{J}{2}\sum^{N-1}_{j=1}(c^\dagger_{j+1}c_j
+c^\dagger_jc_{j+1}) - hN+2h\sum^N_{j=1}c^\dagger_jc_j \nonumber \\
& \ \ \ \ +\frac{J}{2} \left\{ \displaystyle{\prod^N_{k=1}}(1-2c^\dagger_kc_k) \right\} (c^\dagger_1c_N+c^\dagger_N c_1).  
\label{eq.spinless}
\end{align}
We can diagonalize the Hamiltonian (\ref{eq.spinless}) by means of the Fourier transformation.  In the thermodynamic limit ${(N \rightarrow \infty)}$, the ground state energy is given by
\begin{equation}
e=-h+{\frac{1}{2 \pi}}\int^{\kf}_{-\kf}(2h-J \cos q) \rmd q, 
\label{eq.gsenergy}
\end{equation}
where the Fermi point $k_{\rm F}$ is determined by
\begin{equation}
k_{\rm F}=\cos^{-1}(2h/J). \label{eq.fermi}
\end{equation}
Note that $\kf$ goes to zero as ${h \rightarrow \hc \equiv J/2}$.
The EFP (\ref{eq.xyfep}) can be calculated by use of the Fermion operators as
\begin{align}
P(n)&=\langle\prod^n_{j=1}c_jc^\dagger_j\rangle \nonumber \\
&= \det [\langle c_lc^\dagger_m\rangle]^n_{l,m=1} \nonumber \\
&= \det \left[ \frac{1}{2\pi} \int^{2\pi-\kf}_{\kf} 
{\rm e}^{\rmi (l-m) q} \rmd q \right]^n_{l,m=1}, 
\label{eq.toeplitz}
\end{align}
where we have used Wick's theorem.
A kind of the determinant like eq. (\ref{eq.toeplitz}) is referred to as 
a Toeplitz determinant. 

It is known that the transverse correlation function ${\langle
S_r^x S_0^x \rangle}$ for the ${XY}$ model is also written as a slightly different Toeplitz determinant \cite{Lieb61}. It decays algebraically 
as ${(r \rightarrow \infty)}$,
\begin{equation}
 \langle S_r^x S_0^x \rangle \sim 1/\sqrt{r}. 
\end{equation} 
Similarly, 
the longitudinal correlation function decays algebraically 
\begin{equation}
\langle \langle S_r^z S_0^z \rangle \rangle \equiv  \langle S_r^z S_0^z \rangle -  \langle S_r^z \rangle \langle S_0^z \rangle \sim 1/r^2. 
\end{equation}
For further results on these correlation functions, see the refs. 10,12--20. 

As we shall show below, the asymptotic behavior of ${P(n)}$ is remarkably 
different from the two point correlation functions above. 

We remark that there are two other representations for ${P(n)}$. \\
\noindent
(1) Fredholm determinant representation \cite{Colomo92,Essler95n2}: 
\begin{equation}
P(n)=\det \left[I - K_n \right], \label{eq.fredholm}
\end{equation}
where $K_n$ is a linear integral operator acting on functions on 
interval $[-\kf,\kf]$ with a kernel
\begin{align}
K_n(p,q) &= 
\frac{1}{2 \pi} \sum_{j=0}^{n-1} \rme^{\rmi (p-q)j} \nonumber \\
&= \frac{\rme^{\frac{(n-1)(p-q)}{2} \rmi}}{2 \pi} 
\frac{\sin \frac{n}{2}(p-q)}{\sin \frac{1}{2}(p-q)}. \label{eq.kernel}
\end{align}
Or equivalently, 
\begin{align}
P(n)&=\sum_{j=0}^{n} \frac{(-1)^j}{j!} 
\int_{- \kf}^{\kf} {\rm d} q_1 \cdots \int_{-\kf}^{\kf} \rmd q_j \det \left[ K_n (q_l,q_m) \right]_{l,m=1}^{j} \nonumber \\
&= \frac{1}{n!} \int_{\kf}^{2 \pi - \kf} \rmd q_1 
\cdots \int_{\kf}^{2 \pi - \kf} \rmd q_n \det \left[ K_n(q_l,q_m) \right]_{l,m=1}^{n}. \label{eq.fredholm2}
\end{align}
For the derivation of the equalities above, we refer to ref. 24.

Note that the kernel ${K_n(p,q)}$ (\ref{eq.kernel}) can be replaced by 
\begin{equation}
 \frac{\sin \frac{n}{2}(p-q)}{2 \pi \sin \frac{1}{2}(p-q)},
\end{equation}
 due to a property of the determinant \cite{Colomo92}.

\noindent 
(2) Multiple integral representation \cite{Kitanine00,Jimbo96}:
\begin{align}
P(n) &= (-1)^{\frac{n(n-1)}{2}} 2^{\frac{n(n+1)}{2}} \int_{R} 
\frac{\rmd \lambda_1}{2 \pi} \cdots \int_{R} 
\frac{\rmd \lambda_n}{2 \pi} \prod_{a>b} \frac{\sinh 2 (\lambda_a-\lambda_b)}
{\sinh(\lambda_a-\lambda_b -\frac{\pi}{2} i)} \nonumber \\
& \ \ \ \ \ \ \ \ \ \ \ \ \ \ \ \ \times \prod_{j=1}^{n} \frac{\sinh^{j-1}(\lambda_j -\frac{\pi}{4} i) 
\sinh^{n-j}(\lambda_j + \frac{\pi}{4} i)}{\cosh^n 2 \lambda_j}.
\label{eq.multipleintegral}
\end{align}
Here the integration range ${R}$ is defined by
\begin{equation}
\int_{R} \rmd \lambda \equiv \left\{ \begin{array}{ll}
\displaystyle \int_{-\Lambda}^{\Lambda} {\rm d} \lambda 
& \mbox{for ${\kf 
\ge \frac{\pi}{2}}$} \\[2ex]
\displaystyle \left( \int_{-\infty}^{\infty} + \int_{-\infty}^{- \Lambda} + 
\int_{\Lambda}^{\infty} \right) {\rm d} \lambda & \mbox{for  ${\kf
< \frac{\pi}{2}}$} 
\end{array} 
\right. 
\label{eq.range}
\end{equation}
where the endpoint ${\Lambda}$ is related to ${\kf}$ through 
\begin{equation}
\rme^{- 2 \Lambda} = 
-\rmi \frac{\rme^{\rmi \kf} - \rmi}{\rme^{\rmi \kf} + \rmi}.
\label{eq.k-Lambda}
\end{equation}
The integral representation (\ref{eq.multipleintegral}) was first obtained  
in ref. 25,
in the case of ${\Lambda = \infty}$. Recently it 
was confirmed and generalized to finite ${\Lambda}$ case in ref. 6.

We can show that eqs. (\ref{eq.toeplitz}), (\ref{eq.fredholm}) and (\ref{eq.multipleintegral}), are all equivalent \cite{Mehta91, Deift00, Jimbo96}. For the readers' convenience, we give some brief proofs in  Appendix A and Appendix B. 

Note that the Fredholm determinant representation (1) and the multiple integral representation (2) do exist for the general ${XXZ}$ model (\ref{eq.XXZ}) \cite{Essler95n2, Essler95, Jimbo95, Jimbo96, Kitanine00}, while the Toeplitz determinant (\ref{eq.toeplitz}) is very special to the ${XY}$ model. However, in the following, we discuss the asymptotic properties of $P(n)$ for the isotropic ${XY}$ model (\ref{eq.XY}) based on the representation (\ref{eq.toeplitz}). 

Let us first assume that ${h}$ takes some moderate value, $|h|<h_{\rm c}$.  Then from Widom's theorem \cite{Widom71} for the Toeplitz
determinant (\ref{eq.toeplitz}), we immediately obtain
\begin{equation}
\ln P(n)= n^2 \ln \cos \frac{k_{\rm F}}{2}- \frac{1}{4}\ln 
(n \sin{ \frac{k_{\rm F}}{2}}) + 
\frac{1}{12} \ln 2+3\zeta^{'}(-1)+ o(1), \label{eq.asym1}
\end{equation}
i.e.,
\begin{equation}
P(n) \simeq 2^{\frac{1}{12}}{\rm e}^{3\zeta^{'}(-1)} 
\left( \sin \frac{\kf}{2} \right)^{-\frac{1}{4}} n^{-1/4}
\left( \cos \frac{\kf}{2} \right)^{n^2}, \ \ \ \  (n \rightarrow \infty),  \label{eq.asym2}
\end{equation}
where $\zeta(x)$ is Riemann's zeta function. We note that ${\zeta^{'}(-1)
= -0.165421 \cdots}$ and 
\begin{equation}
2^{\frac{1}{12}}\rme^{3\zeta^{'}(-1)} = 0.645002 \cdots.
\end{equation} 
In terms of the external magnetic field $h$, the formula 
(\ref{eq.asym2}) can be expressed as 
\begin{equation}
P(n) \simeq 2^{\frac{1}{12}}\rme^{3\zeta^{'}(-1)}
\left(\frac{1-2h/J}{2}\right)^{-\frac{1}{8}} n^{-\frac{1}{4}}
\left( \frac{1+2h/J}{2} \right)^{\frac{n^2}{2}}, 
\ \ \ \ (n \rightarrow \infty). \label{eq.asym3}
\end{equation}
Thus $P(n)$ for the isotropic ${XY}$ model
(\ref{eq.XY}) decays like a Gaussian.  The speed of Gaussian decay as
well as the prefactor depends on the magnetic field $h$. 
The fact that ${P(n)}$ shows Gaussian decay has been established in ref. 5, 23 and 26. Especially we remark that 
the dependence on the magnetic field ${h}$ in the part of Gaussian decay 
was already obtained in ref. 5. Their method is based on  the
solution of the Riemann-Hilbert problem associated with the Fredholm
determinant (\ref{eq.fredholm}). However, the prefactor
and the part ${n^{-1/4}}$ have not been obtained from the
Riemann-Hilbert approach yet. 

We observe that the prefactor of (\ref{eq.asym1}) diverges as ${h}$ gets
closer to the critical value, i.e., ${h \rightarrow h_{\rm c} = J/2}$.
 There the asymptotic formula (\ref{eq.asym3}) may not be adequate. For
this region, as ${k_{\rm F}}$ is very small, we can expand the
determinant (\ref{eq.toeplitz}) with respect to ${k_{\rm F}}$. Actually we can easily obtain the expansion, 
\begin{align}
P(n) &=1-\frac{\kf}{\pi}n+\frac{\kf^4}{36\pi^2}n^2(n^2-1) - \frac{\kf^6}{675 \pi^2} n^2(n^2-1)(n^2 - \frac{3}{2}) +\cdots,
\label{eq.smallkf}
\end{align}
which may be useful to study the critical behaviors of ${P(n)}$ near 
${h=h_{\rm c}}$.

\begin{figure}[htbp]
\begin{center}
\includegraphics{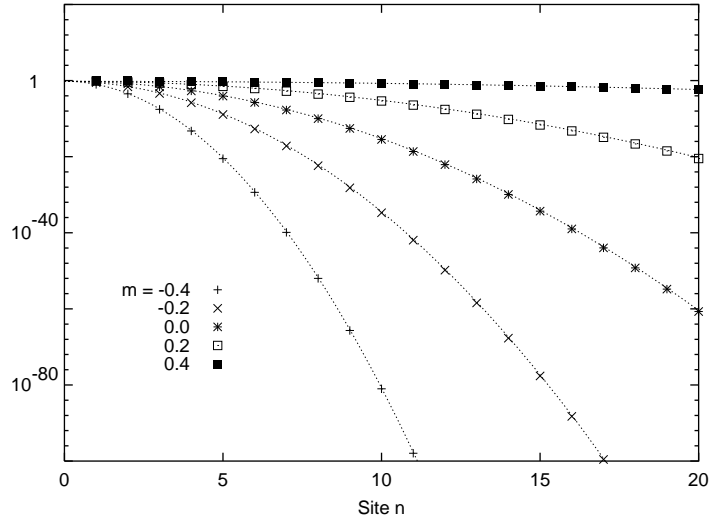}
\end{center}
\caption{$P(n)$ at zero temperature.}
\label{toeplitz} 
\end{figure}

In Fig. \ref{toeplitz}, we plot ${P(n)}$ and the corresponding 
asymptotic function (\ref{eq.asym2}) for several magnetizations ${m
\equiv \langle S_j^z \rangle}$.  
Here note that ${m}$ is connected to ${k_{\rm F}}$ as
\begin{equation}
m = \frac{1}{2} - \frac{\kf}{\pi}.
\end{equation} 
The readers will observe ${P(n)}$ is well fitted by the asymptotic function 
(\ref{eq.asym2}).

We would like to remark that the Toeplitz determinant (\ref{eq.toeplitz}) 
appears in relation to the level spacing distribution of the random
matrix theory \cite{Mehta91}. More precisely, the function
${A_2(\alpha)}$ for  Dyson's circular ensemble of ${n \times n}$ unitary
matrices, is {\it identical} to eq. (\ref{eq.toeplitz}) with the
correspondence ${\alpha = \kf}$. Here ${A_2(\alpha)}$ means the
probability that an interval (of the unit circle) of length ${2 \alpha}$
contains no eigenvalues.  The function ${A_2(\alpha)}$ has been
studied profoundly in the context of the random matrix theory. Then we
can translate all the results obtained for ${A_2(\alpha)}$ to ${P(n)}$.
For example, it is recently shown that ${A_2(\alpha)}$ satisfies an 
integrable non-linear differential equation, which is equivalent to 
Painlev\'{e} VI equation \cite{Tracy94, Deift97}. Hence our ${P(n)}$ 
also satisfies the Painlev\'{e} VI equation \cite{Deift97}. 

\section{Numerical Simulation by DMRG Method}
In this section, we perform numerical simulation so as to confirm 
the validity of the asymptotic form (\ref{eq.asym2}).
We employed the density-matrix renormalization group (DMRG) here
\cite{White92,White93}.The method has an advantage in that it
allows us to treat large system sizes. 
Our algorithm here is standard, and its detail would be found in literatures;
we refer the readers to ref. 30.
Hence, below, we will outline some technical points that are relevant
to our simulation precision:
We implemented the infinite-system method, which is adequate to
study the ground-state properties in thermodynamic limit.
We have repeated hundreds of renormalizations. 
At each renormalization, we remained, at most, two-hundred relevant 
states (bases) for a block; namely, in conventional terminology, we 
set $m=200$.
The density-matrix eigenvalue $\{ w_\alpha \}$ of each remained base
indicates its significance (weight). 
We found $w_{\alpha} > 10^{-10}$.
That is, weights of discarded states are no more than $w_\alpha \sim 10^{-10}$.

However, there are some subtleties in the DMRG calculation when ${S=1/2}$ as in our present case. The number of spins consisting each block increases by one after another through each renormalization. The problem is that the structure of the Hilbert space changes significantly according to the situations whether block contains even number of spins or odd number of spins. (Note that for ${S=1}$, such difficulty does not arise.) Therefore simulation data alternate in turn through renormalizations, even though we repeat hundreds of renormalizations. In this respect, the translational symmetry is broken intrinsically in the scheme. In other words, edge effect remains through renormalizations. In fact, as for ${P(2)}$, for instance, the digit of order ${10^{-3}}$ alternates. We coped with the difficulty by averaging the data over two successive renormalizations. Then we could achieve the precision of ${10^{-4}}$ for ${P(2)}$. The precision of data would be improved further if we use a certain careful extrapolation.  However, in the following, we shall show  that our data reproduces the analytical results rather satisfactorily. 

\begin{figure}[htbp]
\begin{center}
\includegraphics{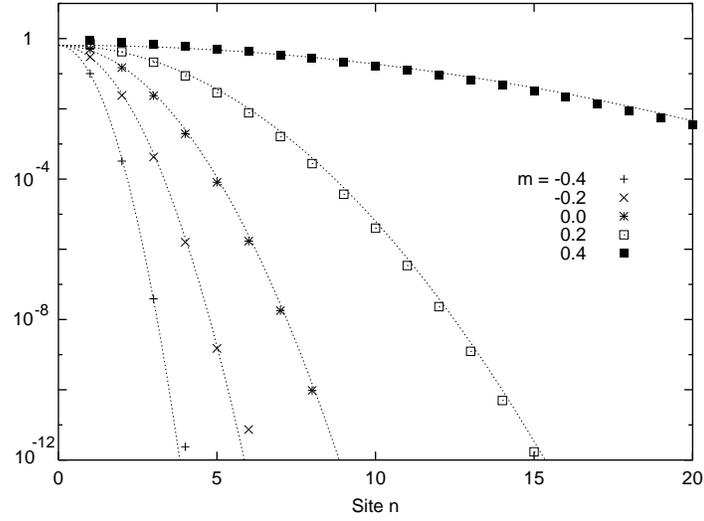}
\end{center}
\caption{Numerical calculation of $P(n)$ by DMRG.}
\label{DMRG} 
\end{figure}

In Fig. \ref{DMRG}, we plot the data for ${P(n)}$ obtained by the DMRG simulation. The dotted lines represent the asymptotic functions (\ref{eq.asym2}). 
For the positive magnetization ${m = 0,0.2,0.4}$, we also list the
explicit numerical data in Table \ref{table1}--\ref{table3}. We find our numerical data by DMRG agree well with the analytical ones in \S 2.   

\begin{table}[htbp]
\begin{center}
\caption{${P(n)}$ for ${\langle S_j^z \rangle =0}$}
\label{table1}
\begin{tabular}{@{\hspace{\tabcolsep}\extracolsep{\fill}}ccc} \hline
    ${n}$ & DMRG  & exact values   
\\ \hline
1     &  ${0.500}$                 &  ${0.5}$          \\
2     &  ${1.486 \times 10^{-1}}$  &  ${1.487 \times 10^{-1}}$  \\ 
3     &  ${2.367 \times 10^{-2}}$  &  ${2.368 \times 10^{-2}}$  \\ 
4     &  ${1.940 \times 10^{-3}}$  &  ${1.945 \times 10^{-3}}$  \\ 
5     &  ${8.096 \times 10^{-5}}$  &  ${8.126 \times 10^{-5}}$  \\ 
6     &  ${1.699 \times 10^{-6}}$  &  ${1.715 \times 10^{-6}}$  \\ 
7     &  ${1.795 \times 10^{-8}}$  &  ${1.823 \times 10^{-8}}$  \\ 
8     &  ${9.559 \times 10^{-11}}$ &  ${9.740 \times 10^{-11}}$ \\ 
9     &  ${4.578 \times 10^{-13}}$ &  ${2.612 \times 10^{-13}}$ \\ 
10    &  ${2.442 \times 10^{-14}}$ &  ${3.514 \times 10^{-14}}$ \\ \hline
\end{tabular}
\end{center}
\end{table}

\begin{table}[htpb]
\begin{center}
\caption{${P(n)}$ for ${ \langle S_j^z \rangle =0.2}$}
\label{table2}
\begin{tabular}{@{\hspace{\tabcolsep}\extracolsep{\fill}}ccc} \hline
$n$ & DMRG  & exact values  
\\ \hline
1     &  ${0.699}$                \ &  ${0.7}$                  \ \\
2     &  ${4.219 \times 10^{-1}}$ \ &  ${4.237 \times 10^{-1}}$ \ \\ 
3     &  ${2.123 \times 10^{-2}}$ \ &  ${2.140 \times 10^{-2}}$ \ \\ 
4     &  ${8.717 \times 10^{-2}}$ \ &  ${8.840 \times 10^{-2}}$ \ \\ 
5     &  ${2.890 \times 10^{-2}}$ \ &  ${2.950 \times 10^{-2}}$ \ \\ 
6     &  ${7.673 \times 10^{-3}}$ \ &  ${7.907 \times 10^{-3}}$ \ \\ 
7     &  ${1.628 \times 10^{-3}}$ \ &  ${1.695 \times 10^{-3}}$ \ \\ 
8     &  ${2.750 \times 10^{-4}}$ \ &  ${2.902 \times 10^{-4}}$ \ \\ 
9     &  ${3.713 \times 10^{-5}}$ \ &  ${3.960 \times 10^{-5}}$ \ \\ 
10    &  ${4.004 \times 10^{-6}}$ \ &  ${4.304 \times 10^{-6}}$ \ \\ \hline
\end{tabular}
\end{center}

\end{table}
\begin{table}[htbp]
\caption{${P(n)}$ for ${ \langle S_j^z \rangle = 0.4}$}
\label{table3}
\begin{center}
\begin{tabular}{@{\hspace{\tabcolsep}\extracolsep{\fill}}ccc} \hline
$n$ & DMRG & exact values \ \ \ 
\\ \hline
1     &  ${0.899}$                \ &  ${0.9}$                  \ \\
2     &  ${7.984 \times 10^{-1}}$ \ &  ${8.003 \times 10^{-1}}$ \ \\ 
3     &  ${6.993 \times 10^{-1}}$ \ &  ${7.019 \times 10^{-1}}$ \ \\ 
4     &  ${6.028 \times 10^{-1}}$ \ &  ${6.061 \times 10^{-1}}$ \ \\ 
5     &  ${5.103 \times 10^{-1}}$ \ &  ${5.146 \times 10^{-1}}$ \ \\ 
6     &  ${4.238 \times 10^{-1}}$ \ &  ${4.290 \times 10^{-1}}$ \ \\ 
7     &  ${3.455 \times 10^{-1}}$ \ &  ${3.508 \times 10^{-1}}$ \ \\ 
8     &  ${2.758 \times 10^{-1}}$ \ &  ${2.811 \times 10^{-1}}$ \ \\ 
9     &  ${2.161 \times 10^{-1}}$ \ &  ${2.206 \times 10^{-1}}$ \ \\ 
10    &  ${1.655 \times 10^{-1}}$ \ &  ${1.695 \times 10^{-1}}$ \ \\ \hline
\end{tabular}
\end{center}
\end{table}
We see that for very small $P(n)<10^{-12}$, simulation data start to deviate widely from the analytical ones. This may be simply due to the numerical round-off error; note that (double precision) real number is stored as sixteen digits in computer and the data of very small ${P(n)}$ are not reliable in principle. For large $h(\sim0.4)$, $P(n)$ decays slowly and our data yield results reliable even for very large distances. Actually, from Table 3, we see that the deviation is less than few percents even for ${n = 10}$.

From the numerical data of ${P(n)}$ for small ${n}$, we can estimate a 
plausible asymptotic form,
\begin{equation}
P(n) \simeq {\rm const} \times a^{-n^2} \ n^{\alpha}. \label{eq.fit}
\end{equation}
In fact, by fitting successive three data for ${P(n)}$ in the form (\ref{eq.fit}), we get the approximate values for ${{\rm const},  a}$ and ${\alpha}$.  We show an example in Table \ref{table4}, where each parameter is obtained from ${P(k), P(k+1)}$ and ${P(k+2)}$. 
\begin{table}[htbp]
\caption{Estimation of the fitting parameters by DMRG: ${ \langle S_j^z \rangle =0}$}
\label{table4}
\begin{center}
\begin{tabular}{@{\hspace{\tabcolsep}\extracolsep{\fill}}cccc}  \hline 
${k}$ &  ${a}$     &  const  &  ${\alpha}$ \\ \hline
1     &  ${1.415}$ & ${0.707}$   &  ${-0.247}$ \\ 
2     &  ${1.415}$ & ${0.709}$   &  ${-0.251}$ \\ 
3     &  ${1.414}$ & ${0.719}$   &  ${-0.271}$ \\ 
4     &  ${1.416}$ & ${0.673}$   &  ${-0.204}$ \\ 
5     &  ${1.415}$ & ${0.723}$   &  ${-0.265}$ \\ \hline
\end{tabular}
\end{center}

\end{table}
The exact parameters in the asymptotic function (\ref{eq.asym2}) are 
\begin{equation}
a = 1.414, \ \ {\rm const} = 0.703, \ \ \alpha = -0.25.
\end{equation}
Thus we could perfectly estimate the speed of Gaussian decay, ${a}$. 
We also got reasonable approximate values for the prefactor and the exponent ${\alpha}$. From these results, we can insist that the DMRG method provides a powerful means to study the EFP at zero temperature for the 1D spin systems. 

\section{EFP at Finite Temperature}
At finite temperature $(T>0)$, the EFP is defined by
\begin{equation}
P(n) = \langle \prod^n_{j=1}(S^z_j+ \frac{1}{2}) \rangle_T \equiv \frac{{\rm Tr} (\rme^{- \calH/T}\prod^n_{j=1}(S^z_j + \frac{1}{2}))}{{\rm Tr}(\rme^{- \calH/T})}.\label{eq.efpTdef}
\end{equation}
For the isotropic ${XY}$ model, we can represent eq. (\ref{eq.efpTdef}) in 
terms of a Toeplitz determinant by use of the finite temperature version of Wick's theorem  (Bloch-de Dominicis theorem) as follows.  
\begin{align}
P(n) &= \det \left[ \langle c_l c_m^\dagger \rangle_T \right]_{l,m=1}^n \nonumber \\
&= \det \left[ \frac{1}{2\pi} \int_{0}^{2\pi} \rme^{\rmi (l-m)q}\rho(q) \rmd q \right]_{l,m=1}^{n}, \label{eq.efpT}
\end{align}
where
\begin{align}
\rho(q) &= \frac{1}{1+ \rme^{- \varepsilon(q)/T}} = 1-\frac{1}{1+ \rme^{\varepsilon(q)/T}}, \nonumber \\
\varepsilon(q) &= 2h-J\cos q.  \label{eq.rhoT}
\end{align}
Note that ${1/(1+\rme^{\varepsilon(q)/T})}$ is nothing but
the Fermi distribution function. 
 Compared with the zero temperature case, we can see that the integration range in each matrix element is replaced as
\begin{equation}
\int_{\kf}^{2\pi-\kf} \rmd q \longrightarrow 
\int_{0}^{2\pi} \rho(q) \rmd q. 
\end{equation}
However, the asymptotic behavior of eq. (\ref{eq.efpT})  
is quite different from eq. (\ref{eq.toeplitz}) . 
In fact, this time we can apply Szeg\"{o}'s theorem to the Toeplitz determinant like (\ref{eq.efpT}) (see Chapter 10 in ref. 31).
The theorem tells us 
\begin{equation}
\ln P(n)=n \rho_0+\sum^\infty_{k=1}k\rho_k \rho_{-k} + o(1), \label{eq.asymT}
\end{equation}
where
\begin{equation}
\ln \rho(q)=\sum^\infty_{k=-\infty} \rho_k \rme^{\rmi k q}.
\end{equation}
In particular, we have
\begin{equation}
\rho_0=-{\frac{1}{2\pi}}\int_{0}^{2 \pi} \ln (1+{\rm e}^{-\varepsilon(q)/T}){\rm d}q,
\end{equation} 
Now recall that the free energy per site for the isotropic ${XY}$ model is given by \cite{Lieb61,Katsura62}
\begin{equation}
f(T,h) = -h-\frac{T}{2\pi} \int_{0}^{2 \pi}\ln(1+{\rm e}^{- \varepsilon(q)/T}) {\rm} {\rm d} q. \label{eq.fenergy}
\end{equation}
Then we can conclude that $P(n)$ decays exponentially at finite temperature as
\begin{equation}
P(n) \simeq c(T,h) \exp \left( -\frac{n}{\xi} \right). \label{eq.asymT2}
\end{equation}
where the inverse of the correlation length ${\xi^{-1}}$ and the prefactor $c(T,h)$ are given by
\begin{align} 
\xi^{-1} &= -\frac{f(T,h)+h}{T} \nonumber \\
        &= \frac{1}{2 \pi} \int_{0}^{2 \pi} \ln \left[1 + 
\exp \left( \frac{J \cos q - 2 h}{T} \right) \right] \rmd q, 
\label{eq.xi} \\
c(T,h) &= \exp \left( \sum^\infty_{k=1} k \rho_k\rho_{-k} \right), \label{eq.prefactor}
\end{align}
respectively. Note a formula 
\begin{equation}
\sum^\infty_{k=1} k \rho_k\rho_{-k} 
= \frac{1}{8} \int_{0}^{2 \pi} \frac{{\rm d}q}{2 \pi} \int_{0}^{2 \pi} \frac{{\rm d} \tilde{q}}{2 \pi}
\left( \frac{\ln \rho(q) - \ln \rho(\tilde{q})}{\sin \frac{1}{2}(q -\tilde{q})} \right)^2, \label{eq.prefactor2}
\end{equation}
which can be found on page 112 of ref. 32. It is useful for numerical evaluation of ${c(T,h)}$ (\ref{eq.prefactor}).

At finite temperature, it is known that the correlation functions such as ${\langle S_r^x S_0^x \rangle}$ and ${\langle \langle S_r^z S_0^z \rangle \rangle}$ also decay exponentially.  The correlation lengths of these correlation functions have been calculated analytically in refs. 15 and 33--35 as
\begin{align}
\xi_{xx}^{-1} &= \frac{1}{2 \pi} \int_{0}^{2 \pi} \rmd q \ln \left( \coth \left| \frac{J \cos q - 2h}{2 T} \right| \right), \label{eq.corrxx} \\
\xi_{zz}^{-1} &= \sinh^{-1} \left( \frac{\pi T - 2ih}{J} \right) 
+ \sinh^{-1} \left( \frac{\pi T + 2ih}{J} \right). 
\label{eq.corrzz}
\end{align}

\begin{figure}[htbp]
\begin{center}
\includegraphics{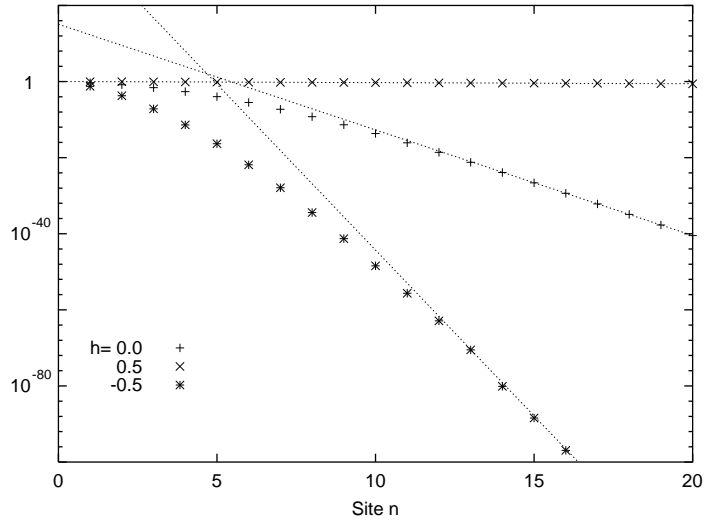}
\end{center}
\caption{$P(n)$ at a low temperature : ${J=1, T=0.05}$.}
\label{lowT} 
\end{figure}

\begin{figure}[htbp]
\begin{center} 
\includegraphics{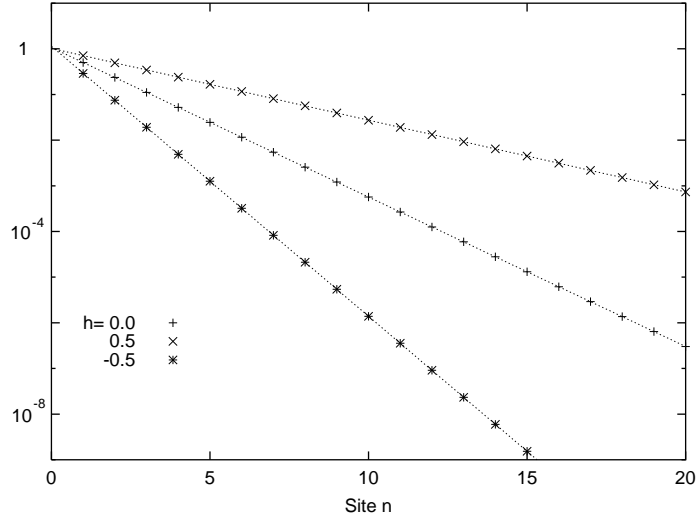}
\end{center}
\caption{$P(n)$ at a high temperature : ${J=1, T=1}$.}
\label{highT} 
\end{figure}

In Fig. \ref{lowT} and Fig. \ref{highT}, we show some plots of ${P(n)}$  calculated from eq. (\ref{eq.efpT}) for ${T=0.05}$ and ${T=1}$. The dotted lines are the asymptotic functions (\ref{eq.asymT2}). 

 We observe the asymptotic behaviors of ${P(n)}$ are actually described by eq. 
(\ref{eq.asymT2}). Particularly in the high temperature case (${T=1}$), even ${P(n)}$ 
for small ${n}$ are well fitted by eq. (\ref{eq.asymT2}).  At low temperature 
(${T=0.05}$), ${P(n)}$ deviates from eq. (\ref{eq.asymT2}) for small ${n}$. At this region, ${P(n)}$ may behave more like Gaussian as in the zero temperature case. 

It is known that, when ${|h| < \hc}$, the low temperature expansion of 
the free-energy (\ref{eq.fenergy}) is given by \cite{Katsura62,Takahashi73},
\begin{equation}
f(T,h) = e - \frac{\pi T^2}{6 v_{\rm s}} + O(T^3), \label{eq.lowTfenergy}
\end{equation}
where ${e}$ is the ground state energy (\ref{eq.gsenergy}) and ${v_{\rm s}}$ 
is the velocity of a low energy excitation
\begin{equation}
v_{\rm s} = J \cos^{-1}\left(2h/J \right).
\end{equation}
From eq. (\ref{eq.lowTfenergy}), we can obtain the low temperature expansion of ${\xi^{-1}}$  as
\begin{align}
\xi^{-1} &= \frac{1}{\pi T} \left\{ J \sin \left[ \cos^{-1} \left( 2h/J \right) \right] - 2h \cos^{-1} \left(2h/J \right)  \right\} 
\nonumber \\
& \ \ \ \ + \frac{\pi T}{6J \cos^{-1}(2h/J)} + O(T^2). 
\label{eq.lowTcorr}
\end{align}
Thus we find the inverse of the correlation length ${\xi^{-1}}$ diverges when ${T \rightarrow 0}$ as long as ${|h| <h_{\rm c}}$. That is, the correlation length ${\xi}$ approaches to zero in the low temperature limit, which signals a crossover of the exponential decay to a stronger (Gaussian) decay in \S 2 and \S 3. This is in a remarkable contrast to the low temperature behaviors of ${\xi_{xx}}$ (\ref{eq.corrxx}) and ${\xi_{zz}}$ (\ref{eq.corrzz}). They behave as ${ \sim 1/T}$ in the low temperature limit, which results from a crossover to a weaker (algebraic) decay.

When the magnetic field is critical, i.e., ${h = \pm \hc}$ \ ${(\hc = J/2)}$, the low temperature expansion of the free-energy takes a differnt form \cite{Takahashi73}, 
\begin{equation}
f(T,h) = - \hc - \frac{1}{\sqrt{2 \pi J}} 
\zeta \left( \frac{3}{2} \right) \left(1 - \frac{1}{\sqrt{2}} \right) T^{3/2} + O(T^2).
\end{equation}
Accordingly, the low temperature behavior of ${\xi^{-1}}$ at ${h = \pm \hc}$ is given by
\begin{equation}
\xi^{-1} = \left\{ \begin{array}{l}
\displaystyle \frac{1}{\sqrt{2 \pi J}} 
\zeta \left( \frac{3}{2} \right) \left(1 - \frac{1}{\sqrt{2}} \right) \sqrt{T} 
+ O(T), \ \ \ \ \ \ \ \ \ \ \ \  \mbox{(for ${h=\hc}$)} \\[2ex]
\displaystyle \frac{2 \hc}{T} + \frac{1}{\sqrt{2 \pi J}} 
\zeta \left( \frac{3}{2} \right) \left(1 - \frac{1}{\sqrt{2}} \right) \sqrt{T} 
+ O(T), \ \ \ \   \mbox{(for ${h=-\hc}$)} \label{eq.lowTcritical}
\end{array} 
\right. 
\end{equation}
respectively. Thus we find particularly that ${\xi^{-1}}$ goes to zero as ${\sim \sqrt{T}}$ when ${h=\hc}$. 

\begin{figure}[htbp]
\begin{center} 
\includegraphics{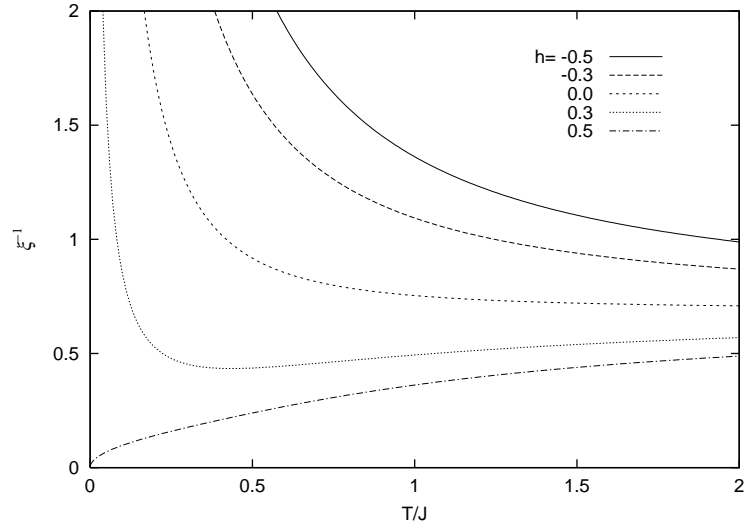}
\end{center}
\caption{Temperature dependence of ${\xi^{-1}}$ : ${J=1}$.}
\label{invcorr} 
\end{figure}

\begin{figure}[htbp]
\begin{center} 
\includegraphics{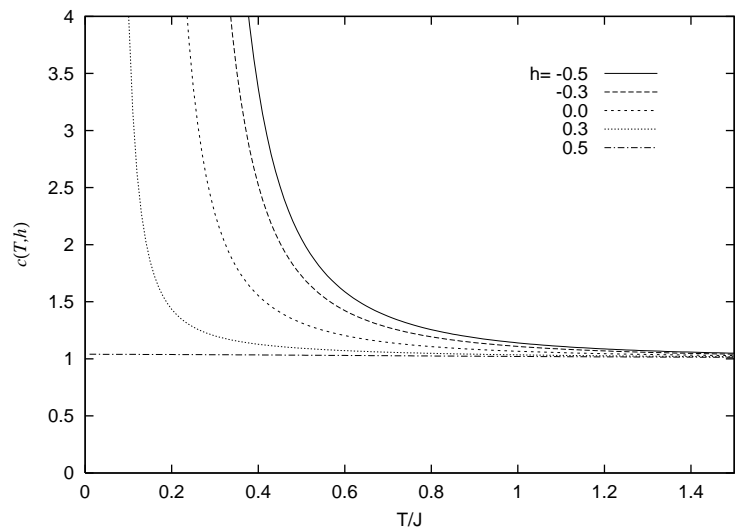}
\end{center}
\caption{Temperature dependence of the prefactor ${c(T,h)}$ : ${J=1}$.}
\label{prefac} 
\end{figure}

In Fig. \ref{invcorr}, we plot the temperature dependences of ${\xi^{-1}}$.  We  observe  the low temperature behaviors of ${\xi^{-1}}$ are actually described by  eqs. (\ref{eq.lowTcorr}) and (\ref{eq.lowTcritical}). 
We also plot the prefactor ${c(T,h)}$ in Fig. \ref{prefac}.  
When ${h < \hc}$, the prefactor ${c(T,h)}$ diverges in the 
low temperature limit.

In the high temperature limit (${T \rightarrow \infty}$), we have
\begin{equation}
P(n) \rightarrow \left( \frac{1}{2} \right)^n, 
\end{equation}
since
\begin{equation}
\xi^{-1} \rightarrow  \ln(2), \ \ c(T,h) \rightarrow 1.
\end{equation}
Alternatively, we can regard the magnetic field $h$ as a function of the temperature and fix the magnetization 
\begin{equation}
m \equiv \langle S_j^z \rangle_T = \frac{1}{2} - \frac{1}{2 \pi} \int_{0}^{2 \pi} \rmd q \frac{1}{1 + \rme^{\varepsilon(q)/T}}. 
\label{eq.magnetization}
\end{equation}
In this case, since
\begin{equation}
\ln \rho(q) \rightarrow \frac{1}{1 + \rme^{- 2h/T}}, \ \ \ \ \ \ \ \ m \rightarrow \frac{1}{2} - \frac{1}{1 + \rme^{2h/T}}, 
\end{equation}
we have
\begin{equation}
P(n) \rightarrow  \left( \frac{1}{2} + m \right)^n = P(1)^n,
\end{equation}
as ${T \rightarrow \infty}$.
\section{Numerical Simulation by QMC Method}
In this section, we perform numerical simulations
in order to confirm the analytical theory for ${P(n)}$ at finite temperatures 
described in the previous section.
Here we employ the quantum Monte-Carlo (QMC) method \cite{Suzuki76}.
For finite-temperature calculations, the QMC method is particularly of use. 
In fact, recently, there have been proposed a 
number of substantial improvements:
In ref. 38,
an algorithm involving infinite Trotter-decomposition number is presented; that is, the continuous-time algorithm.
Employing this algorithm, the authors had demonstrated that
the simulation result is completely free from the Trotter-decomposition error. In addition, with the use of global update algorithm 
postulated in refs. 39--41,
the auto-correlation time is reduced to considerable extent so that
we can avoid wasting Monte-Carlo steps (the so-called critical slowing down).
Hence, Monte-Carlo simulation combined with these techniques would
be promising for studying long-range form of $P(n)$ precisely.

\begin{figure}[htbp]
\begin{center} 
\includegraphics{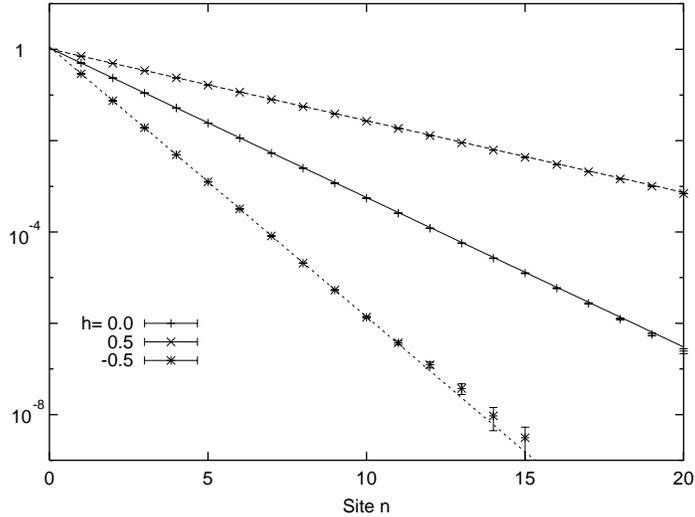}
\end{center}
\caption{$P(n)$ calculated  by QMC method : ${J=1, T=1}$.}
\label{QMC} 
\end{figure}

We treated systems with size $N=128$, and imposed the periodic boundary condition. We performed five-million Monte-Carlo steps initiated by 0.5 million steps for reaching thermal equilibrium. ${P(n)}$ is measured and averaged over the five-million Monte-Carlo steps. The results for ${T=1}$ are plotted in Fig. \ref{QMC}.  We see that the data are governed by the analytical asymptotic function postulated in \S 4 (shown by the dotted lines). We have performed the simulations for several other temperatures and magnetic fields. The obtained data exhibit good coincidence with the analytical ones unless the temperature is exceedingly low ${(T \leq 0.1)}$. In this way we could confirm the validity of the asymptotic formula (\ref{eq.asymT2}) numerically at least for moderate temperatures. Unfortunately when the temperature is very low, the QMC simulation does not work very efficiently so that we could not reproduce the data such as at ${T=0.05}$ yet. 

\section{Summary and Discussion}
In this paper we have studied ${P(n)}$ for the isotropic ${XY}$ model in detail. Especially we have obtained the expression of ${P(n)}$ in terms of Toeplitz determinant. On the basis of the expression, we could calculate the exact values of ${P(n)}$ for small ${n}$ and also get the asymptotic expression as ${n \rightarrow \infty}$. 

At zero temperature, Widom's theorem for Toeplitz determinant gives us the complete asymptotic function directly. It agrees and completes the previous known results on the asymptotics of ${P(n)}$. The asymptotic form of ${P(n)}$ is given by  
\begin{equation}
P(n) \sim a^{-n^2} \ n^{\alpha}, 
\end{equation}
with 
\begin{equation}
a = 1.4142, \ \ \ \ \alpha = -0.25,
\end{equation}
for zero magnetic field. For small ${n}$, we have also calculated 
${P(n)}$ by means of the DMRG method. In general, DMRG is not so efficient to investigate system at criticality such as eq. (\ref{eq.XY}), because any correlation evaluated with the method, in principle, decays exponentially at large distances \cite{Ostlund95}.
However, we have found that as for $P(n)$, which decays like a Gaussian, 
the method yields precise result as well as correct asymptotic form.  We could also estimate the speed of Gaussian decay ${a}$, etc., numerically. The method will be applied to other models in the forthcoming papers.

We could obtain ${P(n)}$ in terms of Toeplitz determinant also at finite temperature. This time from Szeg\"{o}'s theorem, we have shown analytically that ${P(n)}$ decays exponentially as ${n \rightarrow \infty}$,
\begin{equation}
P(n) \sim \exp{\frac{(f+h)n}{T}}. \label{eq.asymformula}
\end{equation}
We have confirmed this formula independently by means of the QMC method. 

It is Boos and Korepin \cite{Boos01} who first obtained the asymptotic formula (\ref{eq.asymformula}) for \emph{any} ${\Delta}$ when ${h=0}$. They derived the formula from an observation of the definition of the thermodynamics, which can be generalized to the finite ${h}$ case straightforwardly \cite{Korepin01}. 
In the forthcoming papers, we will plan to confirm (\ref{eq.asymformula}) for other models by use of the QMC method. It may be also possible to discuss ${P(n)}$ at finite temperature from the point of view of the quantum transfer matrix method \cite{Inoue88, Takahashi91n2,  Kuniba98, Suzuki87,  Koma87, Koma89, JSuzuki90, Takahashi91n1, Kluemper93}. 
\vspace{5mm}
\section*{Acknowledgments}
The authors are very grateful to V. E. Korepin for suggesting us to study the present work and providing us helpful advices.  They also would like to 
thank M. Inoue, Y. Fujii, N. Muramoto and J. Stolze for valuable discussions.
This work is in part supported by Grants-in-Aid for the Scientific Research (B) No. 11440103 from the Ministry of Education, Culture, Sports, Science and Technology, Japan.

\appendix
\section{Equivalence of eq. (\ref{eq.fredholm}) and eq. (\ref{eq.toeplitz})}
Here we will show (\ref{eq.fredholm2}), i.e.,  
\begin{equation}
\frac{1}{n!} \int_{\kf}^{2 \pi - \kf} \frac{\rmd q_1}{2 \pi} \cdots \int_{\kf}^{2 \pi - \kf} \frac{\rmd q_n}{2 \pi} \det \bigg[ \sum_{j=0}^{n-1} \rme^{\rmi (q_l -q_m)j} \bigg]_{l,m=1}^{n}, \label{eq.A-1}
\end{equation}
is transformed to the Toeplitz determinant (\ref{eq.toeplitz}). Actually from 
\begin{align}
 \det \Bigg[ \sum_{j=0}^{n-1} \rme^{\rmi (q_l-q_m)j} \Bigg]_{l,m=1}^{n}
&= \det \left[ {\rm e}^{\rmi q_l j} \right] \det \left[{\rm e}^{- \rmi q_m k} \right] \nonumber \\ 
&= \det \left[ \sum_{l=1}^{n} \rme^{\rmi (j-k) q_l} \right]_{j,k=0}^{n-1}, 
\end{align}
we can see (\ref{eq.A-1}) is equal to  
\begin{equation}
\int_{\kf}^{2 \pi - \kf} \frac{\rmd q_1}{2 \pi} \cdots \int_{\kf}^{2 \pi - \kf} \frac{\rmd q_n}{2 \pi} \det \left[ \rme^{\rmi (j-k) q_k} \right]_{j,k=0}^{n-1} 
=\det \bigg[ \int_{\kf}^{2 \pi - \kf} \frac{\rmd q}{2 \pi} 
\rme^{\rmi(j-k) q} \bigg]_{j,k=0}^{n-1}. 
\end{equation}
The last expression is identical to the Toeplitz determinant (\ref{eq.toeplitz}).

\section{Equivalence of eq. (\ref{eq.multipleintegral}) and eq. (\ref{eq.toeplitz})}
We will show the multiple integral formula (\ref{eq.multipleintegral}) is 
reduced to the Toeplitz determinant (\ref{eq.toeplitz}). Here we assume ${k_{\rm F} \ge \dfrac{\pi}{2}}$ for simplicity. 

From the relation 
\begin{equation}
\sinh(x - \dfrac{\pi}{2} \rmi) = -\rmi \cosh x, 
\end{equation}
 the integral formula (\ref{eq.multipleintegral}) is written as 
\begin{align}
P(n) &= (-\rmi)^{\frac{n(n-1)}{2}} 2^{\frac{n(n+1)}{2}} \int_{-\Lambda}^{\Lambda} \frac{\rmd {\lambda_1}}{2 \pi} \cdots \int_{-\Lambda}^{\Lambda} \frac{\rmd {\lambda_n}}{2 \pi} \nonumber \\
& \ \ \ \ \times \prod_{a>b} 2 \sinh (\lambda_a - \lambda_b)  \prod_{j=1}^{n} \frac{\sinh^{j-1}(\lambda_j -\frac{\pi}{4} \rmi) \sinh^{n-j}(\lambda_j + \frac{\pi}{4} \rmi)}{\cosh^n 2 \lambda_j}. 
\label{eq.B-1}
\end{align}
Furthermore by using the relations 
\begin{align}
\prod_{a>b} 2 \sinh(\lambda_a -\lambda_b) 
&= \prod_{a>b} \left( \rme^{\lambda_a - \lambda_b} - \rme^{\lambda_b-\lambda_a} \right) \nonumber \\
&= \det \left[ \rme^{(2k-n-1) \lambda_j} \right]_{j,k=1}^{n}, 
\end{align}
and 
\begin{equation}
\cosh 2 \lambda_j = 2 \sinh \left(\lambda_j - \frac{\pi}{4} \rmi \right) 
                      \sinh \left(\lambda_j + \frac{\pi}{4} \rmi \right),
\end{equation}
we can find that eq. (\ref{eq.B-1}) is represented by the determinant,
\begin{equation}
P(n) = (-\rmi)^{\frac{n(n-1)}{2}} 2^{\frac{n(n+1)}{2}} \det[I_{jk}]_{j,k=1}^{n}
\end{equation}
where
\begin{equation}
I_{jk} =   \int_{-\Lambda}^{\Lambda} \frac{{\rm d} \lambda}{2 \pi} \frac{2 \rme^{2 k \lambda}}{(\rme^{2 \lambda} - \rmi)^{n - j -1} ({\rm e}^{2 \lambda} + \rmi)^j} .
\end{equation}
Now we introduce a change of integration variable from ${\lambda}$ to ${x}$ defined by 
\begin{equation}
\rme^{2 \lambda} = - \rmi \frac{\rme^{\rmi x} -\rmi}{\rme^{\rmi x} + \rmi}.
\end{equation}
Then after some simple calculations, one can find the determinant representation (\ref{eq.B-1}) becomes
\begin{equation}
P(n) = {(-2 \rmi)}^{\frac{n(1-n)}{2}} \det[\tilde{I}_{jk}]_{j,k=1}^{n}, \label{eq.B-3}
\end{equation}
where
\begin{equation}
\tilde{I}_{jk} =  \int_{k_{\rm F}}^{2\pi - k_{\rm F}} \frac{{\rm d} x}{2 \pi} ({\rm e}^{\rmi x} + \rmi)^{n-k} ({\rm e}^{\rmi x} - \rmi)^{k-1} \rme^{\rmi (j-n) x}. 
\end{equation}
Recall that the relation between ${\Lambda}$ and ${\kf}$ is given in eq. (\ref{eq.k-Lambda}), 
\begin{equation}
\rme^{- 2 \Lambda} = -\rmi \frac{\rme^{\rmi \kf} - \rmi}{\rme^{\rmi \kf} + \rmi}.
\end{equation}

Below we will show eq. (\ref{eq.B-3}) is identical to eq. (\ref{eq.toeplitz}). 
First we subtract the ${(n-1)}$-th column from the ${n}$-th column. Then the integrand of the ${n}$-th column can be replaced as
\begin{equation}
(\rme^{\rmi x} - \rmi)^{n-1} \rme^{\rmi (j-n) x} \rightarrow
- 2i(\rme^{\rmi x} - \rmi)^{n-2}{\rm e}^{\rmi (j-n) x}.
\end{equation} 
Similarly, we can subtract the ${(k-1)}$-th column from the ${k}$-th column subsequently until ${k=2}$. This procedure allows us to replace of the matrix elements for ${k \ge 2}$ as ,   
\begin{equation}
\tilde{I}_{jk} \rightarrow - 2 \rmi \int_{\kf}^{2\pi - \kf} \frac{\rmd x}{2 \pi} (\rme^{\rmi x} + \rmi)^{n-k} (\rme^{\rmi x} - \rmi)^{k-2} \rme^{\rmi (j-n) x}.
\end{equation}
The above transformation yields a factor ${(-2 \rmi)^{n-1}}$, which cancels parts of the prefactor (\ref{eq.B-3}). 

Next we apply the same transformation to the ${n}$-th column again and subsequently to the ${k}$-th column down to ${k=3}$.  This procedure reduces the order of the factor $(\rme^{\rmi x} - \rmi)$ in the integrand by 1 and at the same time generates a factor ${(-2 \rmi)^{n-2}}$. Repeating the similar procedure, we finally arrive at   
\begin{equation}
P(n) = \det \left[ \int_{\kf}^{2\pi - \kf} \frac{\rmd x}{2 \pi} (\rme^{\rmi x} + \rmi)^{n-k} \rme^{\rmi (j-n) x} \right]_{j,k=1}^{n}. 
\label{eq.B-4}
\end{equation}
Expanding ${(\rme^{\rmi x} + \rmi)^{n-k}}$ with respect to ${\rme^{\rmi x}}$,  we can further simplify eq. (\ref{eq.B-4}) as
\begin{align}
P(n) &= \det \left[\int_{\kf}^{2\pi - \kf} \frac{\rmd x}{2 \pi} \rme^{\rmi (n-k)x} \rme^{\rmi (j-n) x} \right]_{j,k=1}^{n}, \nonumber \\
&= \det \left[ \int_{\kf}^{2\pi - \kf}  \frac{\rmd x}{2 \pi} \rme^{\rmi (j-k)x} \right]_{j,k=1}^{n}. 
\label{eq.B-5}
\end{align}
The last expression is nothing but the Toeplitz determinant (\ref{eq.toeplitz}).

\end{document}